# Ring artifacts correction method in x-ray computed tomography based on stripe classification and removal in sinogram images

YANG ZOU,[1,2,6] MEILI QI,[3,6] JIANHUA ZHANG,[4,5] DIFEI ZHANG,[4,5] SHUWEI WANG,[1,2] JIALE ZHANG[1,2], SHENGKUN YAO[1,2,*] AND HUAIDONG JIANG[4,5,*]

[1] *Shandong Provincial Engineering and Technical Center of Light Manipulations & Shandong Provincial Key Laboratory of Optics and Photonic Device, School of Physics and Electronics, Shandong Normal University, Ji'nan 250358, China*
[2] *Joint Research Center of Light Manipulation Science and Photonic Integrated Chip of East China Normal University and Shandong Normal University, East China Normal University, Shanghai 200241, China*
[3] *School of Transportation Civil Engineering, Shandong Jiaotong University, Ji'nan 250357, China*
[4] *School of Physical Science and Technology, ShanghaiTech University, Shanghai 201210, China*
[5] *Center for Transformative Science, ShanghaiTech University, Shanghai 201210, China*
[6] *These authors contributed equally to this work*
*\*Corresponding author: yaoshk@sdnu.edu.cn (S Yao), jianghd@shanghaitech.edu.cn (H Jiang)*

**Abstract:** X-ray computed tomography (CT) is widely utilized in the medical, industrial, and other fields to nondestructively generate three-dimensional structural images of objects. However, CT images are often affected by various artifacts, with ring artifacts being a common occurrence that significantly compromises image quality and subsequent structural interpretation. In this study, a ring artifact correction method based on stripe classification and removal in sinogram images was proposed. The proposed method classifies ring artifacts into single stripes and multiple stripes, which were identified and eliminated using median filtering and multiphase decomposition, respectively. A novel algorithm combining median filtering, polyphase decomposition and median filtering was further developed to eliminate all forms of stripes simultaneously and effectively. The efficacy of the proposed method was validated through both simulated and experimental CT data. The study provides a novel perspective and integrated approach to addressing ring artifacts in X-ray CT. It will be of significant illuminating to a diverse readership, including radiologists, clinical researchers, and industrial scientists.

## 1. Introduction

X-ray computed tomography (CT) is a highly versatile technique that finds applications across various disciplines, including biology, medicine, chemistry [1-4]. The accurate acquisition of CT images is a fundamental prerequisite for the structural interpretation. Nevertheless, the quality of CT images is influenced by a multitude of factors in practice, encompassing ray scattering and strong absorption, data collection errors [5, 6], inaccuracies in reconstruction modelling, and other variables [7-9]. These factors can result in the generation of a variety of artifacts and noises. One of the most commonly observed artifacts is the ring artifacts [10, 11]. It seriously hinders the accurate diagnosis of diseases and the identification of industrial components [12].

Ring artifacts are a common phenomenon in CT imaging, and as a result, there is a substantial body of research dedicated to understanding and mitigating their impact. In general, there are three main methods for the removal of ring artifacts: (1) correction methods based on detector acquisition data, (2) preprocessing methods based on projection data, and (3) post-processing methods based on reconstructed images.

(1) Correction methods based on detector acquisition data mitigate or remove ringing artifacts by resolving inconsistencies in the detector response or modifying the scanning method. In this category, the most classical method is the flat-field correction method [13, 14], which acquires dark-field data without any samples. This data is used to obtain the inconsistency of the detector response, and then it is used to correct the measurement data of the target to eliminate the artifacts. In recent years, further work has been conducted in this area, with Lifton et al. proposing a multi-point, segmented linear flat-field correction in place of the previous two-point flat-field correction. This represents a significant advancement, with the potential to enhance the signal-to-noise ratio of CT data [15]. The approach of Van et al. requires acquiring a series of flat-field images before, during, and after a CT scan to capture the dynamic characteristics of the CT hardware system [16].Ring artifacts preprocessing methods based on the characterization of scintillator variations are alike to the flat field correction methods in terms of the applied theoretical apparatus [17]. Croton et al. proposed a straightforward, pixel-by-pixel detector calibration method utilizing hundreds of data points at each position on the detector. This method can be employed as a standalone calibration or in conjunction with existing ring artifact removal algorithms to enhance image quality further [18]. In the case of processing methods that entail alterations to the scanning method, Zhu et al. employed a hybrid approach involving random detector shifting and data coloring to rectify the projection data [19]. Daniël et al. utilized spiral acquisition to detect the samples, thereby mitigating the impact of ring artifacts [20]. However, this method has high hardware requirements and high operating costs, while the linear response of the detector's pixels must be assumed in the flat-field correction, which can result in the final correction being partially inaccurate.

(2) The second method is the preprocessing method based on projection data. The fundamental concept of this method is that the ring artifacts are represented as stripes in the sinogram of the projection data, which are arranged according to the angle. The stripe artifacts are dealt with in order to reduce the difficulty of detecting and removing them. Kowalski employed a low-pass filter to remove the stripe artifacts. However, this method has the disadvantage of removing some of the high-frequency components, which subsequently decreases the quality of the image [21]. Raven discovered that after Fourier transform, the stripe artifacts in the direction of polar angle can be shifted to the center of the image. Subsequently, a low-pass filter was employed to remove the artifacts [22]. Subsequently, Münch put forth the wavelet-Fourier image stripe removal filter, which employs a combination of wavelet and Fourier transforms to eliminate stripe artifacts while preserving a greater proportion of high-frequency components [23, 24]. Kim et al. proposed a method to remove stripe artifacts from sinograms by estimating the sensitivity of each detector element and equalizing them in sinograms [25]. Lee and colleagues proposed that the filter can be configured to remove ring artifacts through an iterative process [26, 27]. Mäkinen's approach involved transforming the stripes in the sinogram into additive smooth correlation noise through logarithmic transformation. This transformation effectively transformed the task of removing the stripes into the task of removing the noise [28]. Some other preprocessing methods accumulate the information of stripe artifacts combining the bright and dark field images structure and the mean projection image peculiarities [29].

(3) The final method is based on post-processing of reconstructed images. In this category of methods, Axelsson employed local orientation estimation of the image structure and filtering in the reconstructed CT image to identify the ring pattern and thus localize the ring artifacts in the image [30]. Subsequently, Sijbers proposed that the CT image in a Cartesian coordinate system could be transformed to a polar coordinate system, where the artifact region was filtered using a sliding window [31]. Brun further refined Sijbers' method [32]. Subsequently, Wei et al. applied a wavelet transform and Gaussian filtering to CT reconstructed images in polar coordinates to achieve ring artifact removal [33]. Yang et al. employed the concept of decomposition to decompose images in polar coordinates into structural and texture images. They then extracted ring artifacts from the texture images and removed them [34].

In recent years, deep learning techniques have been increasingly employed in image processing, and are being applied to the removal of artifacts in CT images [35-37]. Nevertheless, the majority of extant techniques employ a single methodology for the removal of ring artifacts. In light of the considerable variety observed in ring artifact forms, it is clear that a single approach is insufficient for achieving comprehensive elimination [38]. Note that a considerable proportion of existing methods are not automated: the methods require careful manual search and selection of method parameters, which can be a daunting task for unsophisticated users [39].

In this study, we considered the intricate structural characteristics of the ring artifacts and categorized them as either single or multiple stripes. We propose a method for the correction of ring artifacts based on the classification of stripes, which is both straightforward and rapid to remove all types of ring artifacts. The efficacy of the proposed new method was validated through both simulated and experimental CT data. Furthermore, the range of parameter selection is provided to facilitate the application of our algorithm. This study offers a novel perspective and integrated approach to addressing ring artifacts in X-ray tomography.

## 2. Method

### 2.1 Classification of stripe artifacts

Ring artifacts appear as concentric circles in CT reconstructed slices (Fig. 1b) and as corresponding stripe artifacts in sinogram images (Fig. 1a). The red curve (Fig. 1a) depicts the summation curve, which is calculated by adding the grey values of the columns. In the summation curve, the stripes behave as localized extremes or as a series of abruptly changing values. So, stripes can be classified into two categories on the number of pixels (i.e., the number of abrupt change values). The first category comprises stripes comprising a single or two mutated values, which are henceforth designated as "single stripes" (S1 and S2 in Fig. 1a). The pixels with single stripes have a large intensity difference between them and their neighboring pixels, which can be further classified into two categories. An enhanced single stripe is defined as a bright stripe with a higher intensity than the neighboring pixels (S2 in Fig. 1a). Conversely, a weakened single stripe is defined as a dark stripe with a lower intensity than the neighboring pixels (S1 in Fig. 1a). The second category encompasses stripes comprising three or more pixels, which are designated as "multiple stripes" (S3 in Fig. 1a). S3 in Fig. 1(a) clearly shows multiple consecutive mutations, which appear as a continuous segment in the summed curves and have large mutations from the neighboring normal data.

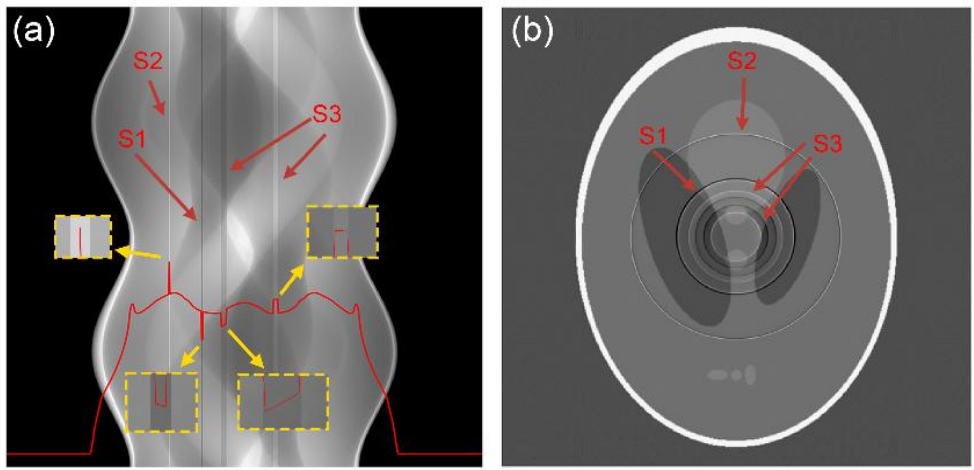

Fig. 1. Ring artifacts in CT image. (a) single stripes (S1 and S2) and multiple stripes (S3) in the sinogram. The red curve is the summation calculated by adding the values of the columns. The stripes can be classified based on the width and intensity. The yellow area presents an enlarged view of the stripe positions within the summation curve. (b) The corresponding concentric ring artifacts of S1, S2 and S3 stripes in the CT slice.

## 2.2 Step 1: removing single stripes formed by a single pixel or two pixels

The ring artifact correction algorithm begins by identifying and eliminating single stripes from the sinogram $P(s, \theta)$. Since the single stripes exhibit abrupt values on the left or right side in the horizontal direction, one-dimensional horizontal median filtering was applied (the size of the filtering kernel is 5), effectively removing single stripes in the filtered image $\bar{P}(s, \theta)$. To locate the single stripes, the difference image $D(s, \theta)$ was generated by subtracting the filtered image from the original sinogram, highlighting the filtered portions. Aggregating the difference image along the column (Eq. 1) differentiates single stripes from other structures.

$$\widetilde{D}(s) = \sum_\theta D(s, \theta) = \sum_\theta \left(P(s, \theta) - \bar{P}(s, \theta)\right), \quad (1)$$

The single stripes are more pronounced in the sum curve $\widetilde{D}(s)$, allowing for the application of two thresholds, $Th$ and $Tl$, to filter the single stripes. Pixels in $\widetilde{D}(s)$ exceeding $Th$ are identified as enhanced single stripes, while those below $Tl$ and non-zero are identified as weakened single stripes. Subsequently, the columns corresponding to all single stripe locations in the sinogram $P(s, \theta)$ are subjected to two-dimensional median filtering to remove the single stripes (we selected a median filter with a window size of 3×5). This method enhances the original algorithm proposed by Yousuf and Asaduzzaman [40] by incorporating a thresholding step that identifies both enhanced and weakened single stripes. Despite its simplicity, median filtering effectively removes single stripes.

Following the removal of single stripes, both the total number and the intensity of remaining stripes in the sinogram were markedly reduced. This improvement in the sinogram facilitated more rapid and efficient processing compared to processing the original sinogram directly. These results underscore the significance and efficacy of the aforementioned procedures.

## 2.3 Step 2: removing multiple stripes formed by multiple pixels

The second step of the ring artifact removal algorithm focuses on eliminating the remaining multiple stripes (Fig. 3). Here, we apply the concept of polyphase decomposition to dissect the sum curve $y(s)$, which aggregates all values in each column of the sinogram (Eq. 2), into $L$ individual curves, where $L$ corresponds to the number of layers in the polyphase decomposition (Eq. 3).

$$y(s) = \sum_\theta \widetilde{P}(s, \theta), \quad (2)$$

$$y_k(s) = y(sL + k - 1), \quad (3)$$

where $1 \leq k \leq L$. From Eq. 3, polyphase decomposition involves sampling the sum curve at intervals of $L$, yielding $L$ subcurves. In the sum curve, multiple stripes appear as a series of abrupt changes in value. After polyphase decomposition, these consecutive abrupt changes are converted into local extrema within the subcurves. This occurs because adjacent values are distributed across different subcurves during the decomposition process, causing the originally continuous abrupt changes to appear as isolated extreme points in the subcurves. The locations of these local extrema correspond to the positions of the single stripes. Determining the positions of all local extremes on $y_1(s), y_2(s), \ldots, y_L(s)$ allows us to obtain the positions of all single stripes, and thus the positions of all multiple stripes. Subsequently, a high-pass filter is applied to $y_k(s)$, $y_h(s) = 2y_k(s) - y_k(s-1) - y_k(s+1)$, which will result in the location of the $y_k(s)$ extreme producing an extreme at the same location of $y_h(s)$, while the sign of both sides is opposite to that of the extreme. This provides the extremum determination condition. If the local extremum is at position a of $y_k(s)$, then $y_h(a)$ is the extremum, with $y_h(a-1)$ and $y_h(a+1)$ having opposite signs to $y_h(a)$, as show in Fig. 2.

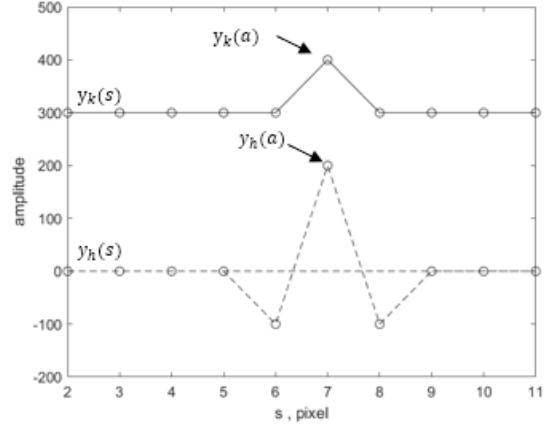

Fig. 2. Processing data using high-pass filtering.

Furthermore, in order to prevent the misclassification of structures as stripes, it is necessary to define a threshold value, $T$, which serves to differentiate between these two categories based on the intensity of the extremes. This can be expressed as follows: $|2y_h(s) - y_h(s-1) - y_h(s+1)| \geq T$. Once the extreme locations of $y_k(s)$ have been identified, the locations of the multiple stripes in $y(s)$ can be determined in accordance with Eq. 3. Thereafter, the defective data can be replaced by linear interpolation. To circumvent the presence of defects on both sides of the defective data, the data employed for interpolation is selected from $y_k(s)$ and utilized to enhance the restoration outcomes.

The concept of applying polyphase decomposition to sinograms was initially proposed by Anas and Lee[10]. Their original algorithm directly applies polyphase decomposition to the sinogram and uses iterative processing to enhance the effectiveness of the treatment. However, in our method, the number of stripes in the sinogram is significantly reduced after the first step. Subsequently, the second step further improves the stripe removal by processing the sinogram, which results in better performance compared to directly applying polyphase decomposition to the original sinogram. Moreover, this approach eliminates the need for iterative processing. The proposed method leads to a more computationally efficient algorithm and avoids the challenge of selecting an appropriate number of iterations. We have improved the removal of ring artifacts by traversing the entire image. The workflow of the second step of the algorithm is illustrated in Fig 3.

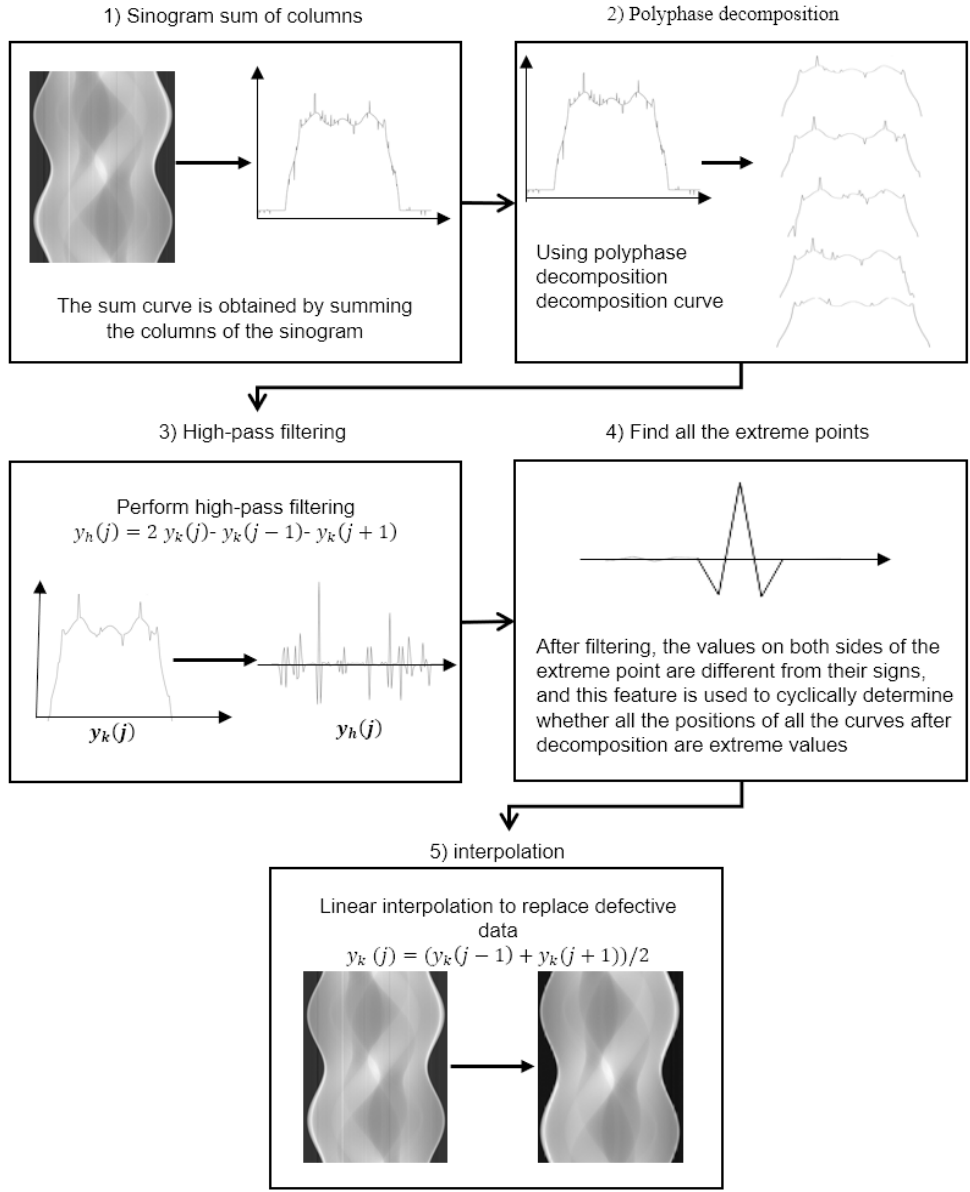

Fig. 3. The multiple stripe removal based on polyphase decomposition. (1) Summing the columns of a sinogram. (2) Decomposing the sum curve using polyphase decomposition. (3) Apply high-pass filtering to the decomposed curve. (4) Find all eligible extreme points. (5) Apply linear interpolation to the columns where the extreme points are located.

### 2.4 Step 3: removing residual stripes

Finally, the third step of our ring artifact removal algorithm removes the additional stripes generated in the second step and any residual stripes not previously eliminated. This is necessary because polyphase decomposition not only decomposes multiple stripes but also leads to structural decomposition, which can generate further stripes. If the number of polyphase decomposition layers, denoted by $L$, is insufficient, the removal of multiple stripes will be incomplete, leaving residual stripes. However, these additional and residual stripes are predominantly single stripes, which can be effectively removed using the first step of our algorithm. The window size of the filter is the same as that in the first step. Thus, our algorithm is structured into three distinct phases: an initial median filtering step, followed by polyphase decomposition, and concluding with another median filtering step.

## 3. Experimental results

### 3.1 Simulated ring artifact removal results

A simulation validation was conducted to ascertain the efficacy of the proposed methodology. The simulation data were obtained using a Shepp-Logan model with a size of 512 by 512. The corresponding sinogram was obtained by radon transform at 360 angles with a range of 0-180°, resulting in a size of 360 by 729. Initially, a combination of Poisson and Gaussian noise was superimposed on the projection data [41, 42]. Subsequently, stripes were introduced by randomly selecting one fifth of the columns in the sinogram and multiplying them by random numbers according to a normal distribution with a standard deviation of 0.05 [43].

We performed the ring artifacts correction by using four methods: median filtering (Algorithm 1), polyphase decomposition (Algorithm 2), wavelet-Fourier transform (Algorithm 3), the combined algorithm presented in this paper (Algorithm 4). Algorithm 1 is a median filtering technique proposed by Yousuf et al, which demonstrates efficacy in removing enhanced single stripes, yet it is ineffectual in removing weakened single stripes or multiple stripes (Fig. 4h). Algorithm 2 is the polyphase decomposition method of Anas et al, which employs a thresholding process at each iteration to filter the stripes. This approach proved effective for removing the majority of the stripes, though it results in the destruction of structures, leading to the formation of more intense stripes (Fig. 4c). Algorithm 3 is based on a wavelet-Fourier transform by Munch et al, which employs frequency domain information in the stripe removal process. The removal of the stripes results in a compromise to the structural integrity (Fig. 4d). The combined algorithm presented in this paper, Algorithm 4, represents an enhanced method of the original methods, capitalizing on their respective strengths and removes stripes while reducing the damage to the structure (Fig. 4k). The simulation demonstrated the effective of proposed method in correction of ring artifacts.

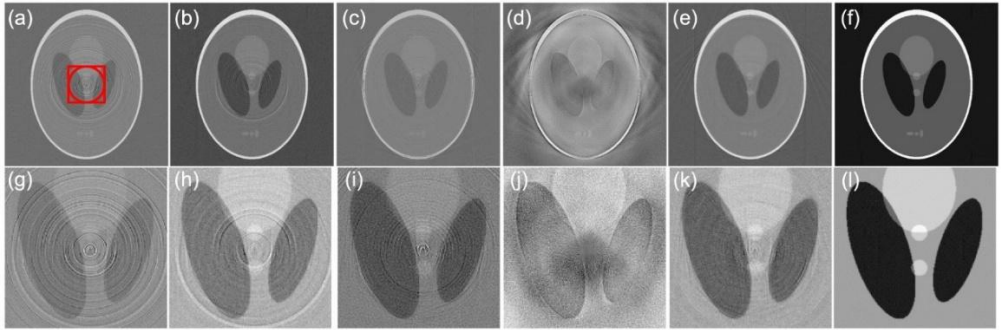

Fig. 4. Simulation of ring artifacts correction. (a) The uncorrected image with ring artifacts. (b) The corrected image by median filtering (Algorithm 1). (c) The corrected image by the polyphase decomposition (Algorithm 2). (d) The corrected image by wavelet-Fourier transform (Algorithm 3). (e) The corrected image by the combined algorithm presented in this paper (Algorithm 4). (f) The reference images. The images in second row are zoomed-in views of the center of the image in the first row.

### 3.2 Experimental data removal results

To further ascertain the efficacy of the proposed methodology, we conducted an evaluation using experimental data, a tomographic slice of an insects, which is performed in synchrotron X-ray tomography at Shanghai Synchrotron Radiation Facility (SSRF).

X-ray images of the yellow mealworm Tenebrio molitor (Fig. 5) were obtained using the BL13W1 beamline at the Shanghai Synchrotron Radiation Facility (SSRF). The experimental setup is schematically illustrated in Fig. 6. A 12 keV X-ray beam, monochromatized by a double Si(111) crystal system, was directed through the sample, capturing absorption details. The transmitted beam was then imaged by a CCD detector with a resolution of 2048 × 2048 pixels (9 × 9 μm per pixel). The specimen was mounted on a precisely calibrated rotary stage parallel to the CCD camera, enabling rotation through 180° with 600 projections captured at an exposure time of 0.055 seconds each. Supplementary white field images—captured without the sample present—were recorded at ninety angles, and five dark field images were taken with no light reaching the detector post-acquisition. The tilted projection series was reconstructed into a three-dimensional representation using filtered back-projection (FBP) algorithms, applying a Shepp-Logan filter for enhanced noise suppression compared to the standard ramp filter.

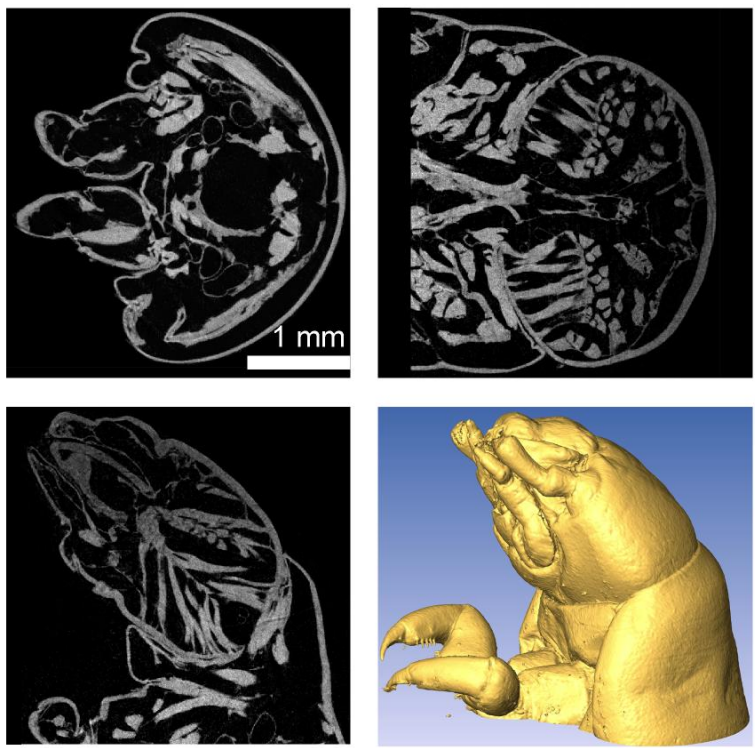

Fig. 5. 3D morphology of the yellow mealworm Tenebrio molitor and three orthogonal slices.

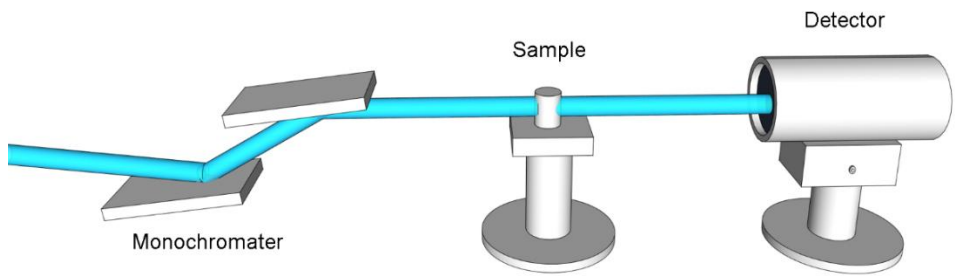

Fig. 6. Schematic layout of the CT setup with absorption contrast.

A CT slice exhibiting ring artifacts (Fig. 7a and 7f) was selected for processing using the four aforementioned algorithms. The application of Algorithms 1 and 2 (Fig. 7b and 7c, and 7g and 7h, respectively) resulted in the destruction of the image structure and the generation of a greater number of artifacts. Some ring artifacts remain in the upper left region of the rotation center in Fig. 7(g), and the ring artifacts in Fig. 7(h) have not been fully removed. Algorithm 3 was effective in eliminating the ring artifacts, although some structural details were lost. The Algorithm 4 (Fig. 7e and 7j) proposed in the study was successful in correcting the ring artifacts situated proximate to the center of rotation. Furthermore, the image noise was reduced and the damage to the image's structural integrity was mitigated, thereby enhancing the image quality. The experimental X-ray CT data demonstrated the efficacy of the proposed method in the correction of ring artifacts.

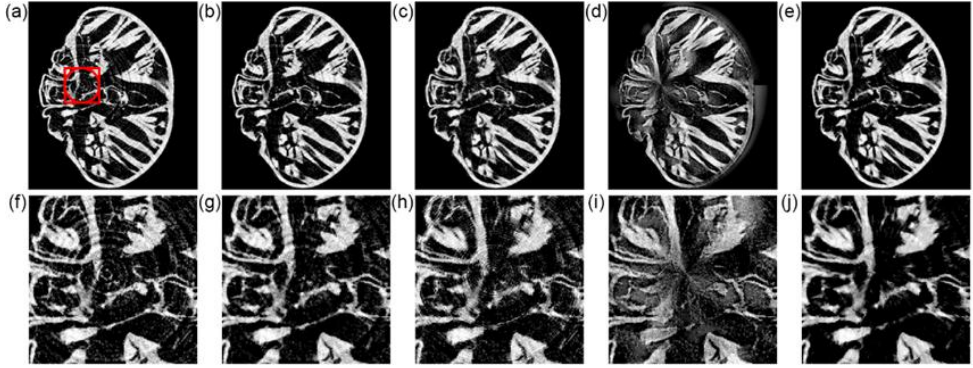

Fig. 7. Ring artifacts correction in experimental X-ray CT data. (a) The uncorrected image; (b) The corrected image by algorithm 1. (c) The corrected image by algorithm 2. (d) The corrected image by algorithm 3. (e) The corrected image by algorithm 4. The images in second row are zoomed-in views of the center of the image in the first cow.

*3.3 Numerical evaluations*

In addition to visual assessment, the effectiveness of the ring artifact correction methods was quantitatively evaluated using the signal-to-noise ratio (SNR) and the ring artifact suppression percentage (RASP) [18, 39]. For the simulated data, the red circular region in Fig. 4(a) was selected for RASP calculation, while the red rectangular region was used for SNR calculation. For the experimental data, the red region in Fig. 7(a) was chosen for metric calculation. The formula for RASP is as follows:

$$RASP = \left(1 - \frac{\sigma_c + C_1}{\sigma_u + C_2}\right) \times 100\% = \left(1 - \frac{\sqrt{\frac{1}{N}\sum_{j=1}^{N}(x_{c,j} - \mu_{c,j})^2} + C_1}{\sqrt{\frac{1}{N}\sum_{j=1}^{N}(x_{u,j} - \mu_{u,j})^2} + C_2}\right) \times 100\%. \quad (4)$$

Here the parameters $c$ and $u$ refer to the corrected and uncorrected images, respectively. $N$ is the total number of radial pixels, $x_j$ is the mean of all pixels in a radial direction, determined by finding a mean value over a radial profile, and $\mu_j$ is the mean value of pixels in a radial direction, determined by smoothing the radial profile using a median filter. So, here, $\sigma_c$ and $\sigma_u$ denote the standard deviation of the radial pixel values of the corrected and uncorrected images, respectively. The constant $C_1$ and $C_2$ are introduced to prevent the denominator from being too close to zero (we set $C_1 = C_2 = 0.01$). In order to facilitate the processing of the data, we transform the reconstructed rectangular coordinate image into a polar coordinate image. This transformation allows the data on a radial contour in the rectangular coordinate image to become the data on a column of pixels in the polar coordinate image. Consequently, the data is calculated and processed in the polar coordinate domain. The RASP of the region of interest can be calculated depending on the region selected.

Table 1 shows that Algorithm 3 achieved the highest RASP value in the simulated data, indicating the most effective removal of ring structures. Algorithm 4 followed with a high RASP value and the highest SNR, demonstrating superior image quality. In the experimental data, Algorithm 4 achieved the highest RASP value, effectively removing ring artifacts while preserving structural details.

**Table 1. The quantitative assessment of different methods in removing ring artifacts using simulation and experimental data**

|  |  | RASP | SNR |
|---|---|---|---|
| Simulation data | Without algorithm | 0 | 3.570 |
|  | Algorithm 1 (Median filtering) | 22.112 | 3.478 |
|  | Algorithm 2 (Polyphase decomposition) | 8.795 | 6.061 |
|  | Algorithm 3 (Wavelet-Fourier transform) | 66.110 | 1.950 |
|  | Algorithm 4 (Combined) | 52.981 | 7.234 |
| Experimental data | Algorithm 1 (Median filtering) | 18.332 | 15.732 |
|  | Algorithm 2 (Polyphase decomposition) | -22.705 | 11.970 |
|  | Algorithm 3 (Wavelet-Fourier transform) | 27.241 | 3.305 |
|  | Algorithm 4 (Combined) | 34.527 | 12.123 |

## 4. Discussion

To facilitate the use of our method, we have displayed the sinograms after each step of the algorithm (Fig 9) and discussed the impact of algorithm parameters on stripe removal. Figure 9c shows the sinogram with ring artifacts. After applying the first step of our algorithm, the results are shown in Fig 9e and 9i. When the number of layers in polyphase decomposition is too large (L=6 in this case), the results of the second and third steps are shown in Fig 9f and 9g, respectively. It can be observed that the median filtering in the first step significantly reduces the number and intensity of artifacts in the sinogram. However, a large value of L leads to the decomposition of structures, especially edge structures (indicated by the blue arrow in Fig 9f). The filtering in the third step mitigates the severity of structure degradation (indicated by the blue arrow in Fig 9g). When the number of layers in multiphase decomposition is too small (L=3 in this case), the results of the second and third steps are shown in Fig 9j and 9k, with the blue area being the magnified part. It is evident that the second step of the algorithm decomposes the multiple stripes in Fig 9i into single stripes (Fig 9j), and the filtering in the third step effectively removes these residual stripes. Therefore, when selecting algorithm parameters, the parameters in the first step do not need to be strictly chosen, as subsequent steps will continue the processing. The number of layers L in the second step should be chosen between 3 and 6; a larger value may lead to structure destruction and the generation of additional stripes. In the third step, since there are very few stripes left in the sinogram, the parameter selection should be more stringent to effectively remove the remaining stripes.

In order to verify the effect of threshold selection in median filtering on the removal of ring artifacts, we chose to apply median filtering alone to noise-free simulated data. We chose the interval [0.01, 2] when selecting Th at 0.01 intervals, i.e., Th selects 200 data; and the interval [0.0001, 0.01] when selecting Tl at 0.0001 intervals, i.e., Tl selects 100 data (Fig. 8a). This is due to the fact that in algorithm, the two extreme cases of using median filtering are filtering all the images and no filtering, respectively. So, in the selection, the minimum value of Th is 0.01 and the maximum value of Tl is 0.01, which corresponds to the case of filtering all of the image; in the case where Th is selected 2 and Tl is selected 0.0001, there is very little processing of the image, which can be approximated to correspond to the case of no processing of the image. In order to verify the effect of the number of polyphase decomposition layers L and the threshold T on the removal of ring artifacts, we choose the maximum number of polyphase decomposition layers to be 5. In the study of the threshold T for the determination of the extreme value, we choose T to be in the range of [0.001,1], and the data are selected every 0.001, this is due to the fact that the number of column pixels processed is already much larger than the number of defective column pixels when T equals to 0.001; and only a few obvious stripes in the image can be processed when T equals to 1, which is already unsatisfactory for the removal of ring artifacts from the whole image.

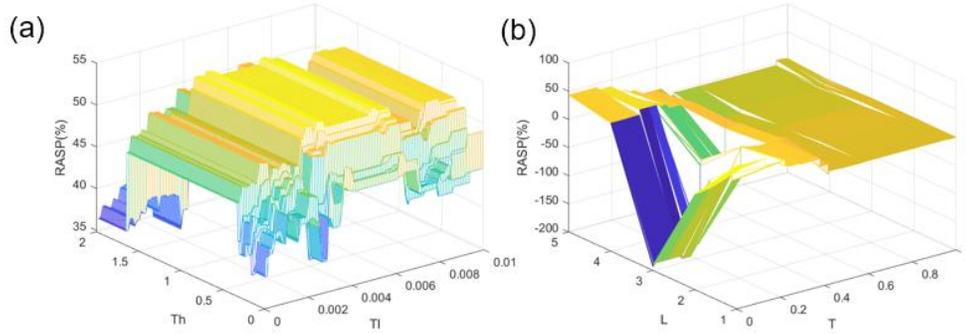

Fig. 8. (a)The variation of RASP with the threshold Th and Tl. (b)The variation of RASP with the threshold T and the number of polyphase decompositions L.

As illustrated in Fig. 8(b), the maximum value of L is 5. This is due to the fact that the maximum number of pixels in the additional multiple stripe that has been incorporated into the image is 5 pixels. Consequently, when the number of polyphase decomposition layers is 5, the method is capable of removing all stripes; however, an increase in the number of polyphase decomposition layers will not result in the desired outcome. Instead, it will have a detrimental impact on the removal of ring artifacts, ultimately destroying the original structure of the image and reducing its quality.

Based on the assessment of the results from the simulation data and RASP, we believe that the algorithm performs well when the Tl value is in the range of [0.003, 0.007], the Th value is in the range of [0.4, 1.6], and the T value is in the range of [0.001, 0.15]. In the experimental data, we investigated the performance of Algorithm 4 in removing ring artifacts and preserving structures by varying the parameter values. The results were consistent with the simulated data. Specifically, the parameter Th in the first and third steps of the algorithm significantly influenced the outcome. A smaller value of Th led to better artifact removal. Therefore, Th is chosen in [0.4, 1.6], Tl has less influence and can also be chosen in [0.003, 0.007]. Additionally, the number of layers L in the polyphase decomposition step of the algorithm had a substantial impact. A larger L resulted in more structural decomposition within the image. However, when $L = 3$, the effectiveness is enhanced, as the remaining fringes after decomposition are precisely single fringes, which can be effectively removed in the third step.

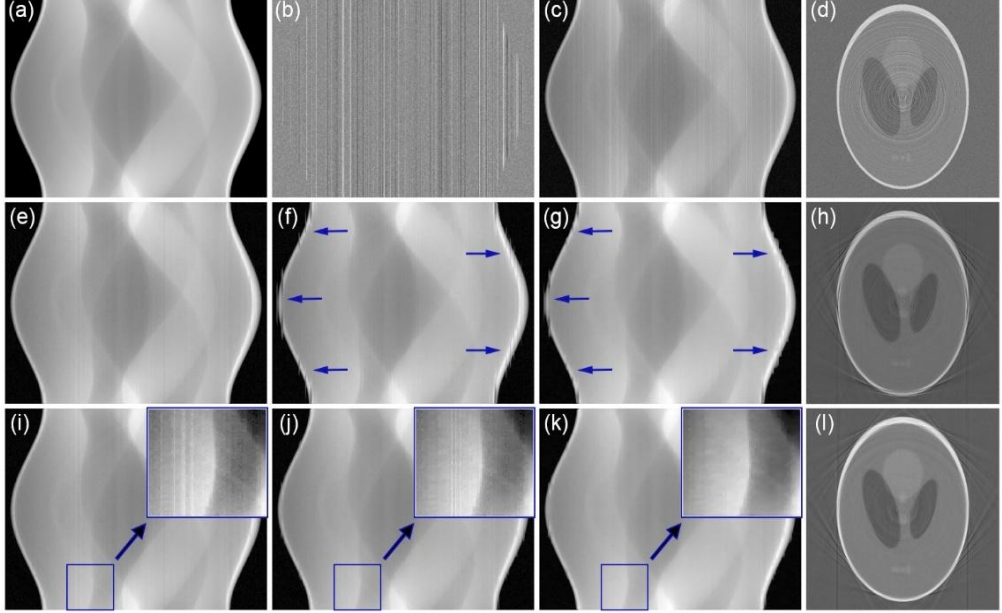

Fig. 9 The intermediate processes of applying the algorithm to the sinogram are as follows: (a) The original sinogram without artifacts. (b) The generated ring artifact component. (c) The sinogram with ring artifacts. (d) The CT reconstruction image with ring artifacts. (e)-(h) The intermediate sinograms and CT reconstruction images of the algorithm when L=6. (i)-(l) The intermediate sinograms and CT reconstruction images of the algorithm when L=3. The blue areas represent the magnified regions.

The results of both simulations and experiments demonstrate that the proposed ring artifact correction method, which is based on stripe classification, is an effective means of removing ring artifacts and restoring the structures. It is important to note that incorrect parameter settings in the algorithm can result in the removal of stripes in an unclean manner or the generation of additional stripes. The selection of Th and Tl is based upon the standard deviation of the stripes; when selecting T and L, it is necessary to normalize the image, thereby ensuring its applicability to the majority of datasets.

The method described herein is a combined removal method that is designed to remove ring artifacts. In comparison to alternative techniques that employ a single method for the removal of ring artifacts, our method demonstrates a broader scope of applications and a reduced propensity for structural damage and the generation of ring artifacts. Nevertheless, the efficacy of the method is limited in the case of weak ring artifacts that are not present across the entire ring. In this instance, the presence of the ring artifacts has no impact on the interpretation of the image. However, our method tends to cause disruption to ring-like structures, while it preserves other areas relatively well. Additionally, the third step of our algorithm only mitigates the damage to edge structures rather than effectively restoring them. This limitation is attributed to the interpolation algorithm employed. With corresponding modifications, the method may be employed as a potential solution for the removal of stripe artifacts in a more expansive range of images. Furthermore, the method can be applied to the ring artifacts removal in CT-reconstructed slices.

## 5. Conclusion

In conclusion, we considered the intricate structural characteristics and categorized ring artifacts as either single or multiple stripes in sinogram and proposed a novel algorithm combining median filtering, polyphase decomposition, and median filtering to effectively eliminate all forms of stripes simultaneously. The efficacy of the proposed method was successfully validated through both simulated and experimental CT data. The study provides a novel perspective and integrated approach to addressing ring artifacts in X-ray CT. This study introduces a novel perspective and presents a comprehensive approach to addressing the issue of ring artifacts in X-ray CT, providing valuable insights to a diverse audience.

**Acknowledgements**

National Natural Science Foundation of China (12004227), Young Talent of Lifting Engineering for Science and Technology in Shandong, China (SDAST2021qt05), Shandong Provincial Natural Science Foundation (ZR2020QA076 and ZR2020QE070). We thank the staffs of beamline 13HB at Shanghai Synchrotron Radiation Facility for assistance with data acquisition.

**Conflict of Interest**

The authors declare no conflict of interest.

**Data availability**

Data will be made available on reasonable request.